\documentclass[5p,twocolumn,times]{elsarticle}

\usepackage{graphicx}
\usepackage{amsmath}   
\begin{document}

\begin{frontmatter}

\title{The Archimedes Experiment}

\author[add2]{Enrico~Calloni}
\author[add3]{S.~Caprara}
\author[add2]{Martina~De~Laurentis}
\author[add2]{Giampiero~Esposito}
\author[add3]{M.~Grilli}
\author[add3]{Ettore~Majorana}
\author[add2]{G.P.~Pepe}
\author[add3]{S.~Petrarca}
\author[add3]{Paola~Puppo}
\author[add3]{P.~Rapagnani}
\author[add3]{Fulvio~Ricci}
\author[add2]{Luigi~Rosa}
\author[add4]{Carlo~Rovelli}
\author[add1]{P.~Ruggi}
\author[add3]{N.L.~Saini}
\author[add2]{Cosimo~Stornaiolo}
\author[add2]{Francesco~Tafuri}

\address[add1]{European Gravitational Observatory (EGO), Cascina (Pisa) }
\address[add2]{University of Napoli Federico II and INFN Napoli}
\address[add3]{University of Roma Sapienza and INFN Roma}
\address[add4]{University of Aix-Marseille}

\begin{abstract}
Archimedes is an INFN-funded pathfinder experiment aimed at verifying the feasibility of measuring the interaction 
of vacuum fluctuations with gravity. The final experiment will measure the force exerted by the gravitational field 
on a Casimir cavity whose vacuum energy is modulated with a superconductive transition, by using a balance as a small 
force detector. Archimedes is two-year project devoted to test the most critical experimental aspects, in particular 
the balance resonance frequency and quality factor, the thermal modulation efficiency and the superconductive sample realization.
\end{abstract}


\end{frontmatter}

\section{Introduction}
The cosmological constant problem belongs to the main fundamental unsolved questions 
which already puzzle the modern physics. 
Does the enormous energy of the vacuum fluctuations foreseen by quantum mechanics contribute to gravity? 
If it does, the vacuum energy density suggested by the quantum field theory is enormously larger than the value 
constrained from General Relativity, by considering the radius of our universe and its accelerated expansion. 
At present, although there is a long list of detailed and important theoretical works \cite{Weinberg,DeWitt,Pad} 
on this argument, there is no agreement on the real contribution of vacuum fluctuations to gravity on the theoretical side. 
Moreover no experiment has been performed to finally verify or discard this assumption so far. 
Taking into account the present status of the small forces measurement techniques and of the technologies of the high 
temperature superconductors, an experimental device to make such a measurement is on the designing path \cite{Pos}.
\begin{figure}
\centering
\includegraphics[width=0.5\linewidth]{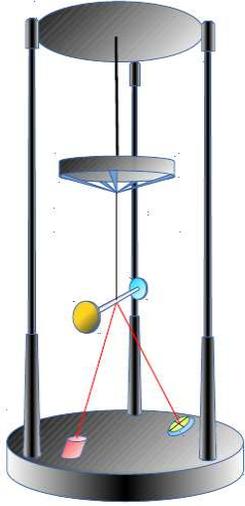}
\caption{Scheme of the balance suspended to the inverted pendulum.}
\label{fig:Balance}
\end{figure}
\section{The weight of the vacuum}
The idea is to weigh the vacuum energy stored in a rigid Casimir cavity formed by parallel conductive plates. The Casimir 
effect is a macroscopic manifestation of vacuum fluctuations. When a Casimir cavity is formed only some modes can resonate 
inside it. The ones which do not satisfy specific boundary conditions are expelled and the total  vacuum energy changes. 
If the vacuum energy does interact with gravity, a force directed upwards acts on the cavity and can be interpreted as the 
lack of weight of the modes expelled by it, in similarity with the Archimedes buoyancy of fluid. The force is equal to \cite{Calloni2}:

\begin{equation}
\vec{F}=-\frac{\left| E_{C}\right|}{c^2} \vec{g}
=-{A} \frac {\pi^2 \hbar}{720~a^3} 
\frac {\vec{g}}{ c}
\end{equation}
where $E_C$ is the vacuum energy stored in the cavity, $a$ is the gap between the plates and $A$ is their surface, $c$ is 
the speed of light and $\vec{g}$ is the earth gravitational acceleration directed
downwards.  

The measurement strategy is to modulate the reflectivity of the plates in such a way as to periodically expel the vacuum energy 
from the Casimir cavity and consequently modulate its weight.
A possible way to modulate the reflectivity is by performing a superconducting transition of the plates: when the plates become 
superconductive the reflectivity changes and so does the vacuum energy contained between the plates \cite{Calloni}. 
The variation of Casimir energy could be particularly relevant in case of type II high  layered $T_c$  superconductors, 
like cuprates which behave as natural multi Casimir-cavities. The advantage is due to the fact that in normal state the 
planes (that will become superconducting) are very poorly conductive. 
Even though a complete study is still lacking, and it is one of the problems to be faced inside the Archimedes project, 
order of magnitude expectations suggest that the amplitude $F$ of the modulated force is of $F\sim10^{-16} N$. 

\begin{figure}
\centering
\includegraphics[width=1.0\linewidth]{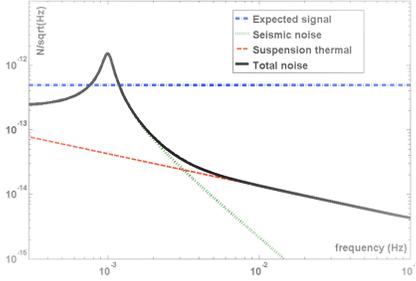}
\caption{Expected signal and main noises. The detection bandwidth is within the resonance of the balance, limited by the suspension 
thermal (dashed line) and seismic noise (dotted line)}
\label{fig:noise}
\end{figure}

\section{The Experiment}
The experiment aim is to measure forces of $F\sim10^{-16} N$. Recently \cite{Calloni2} it has been shown that both 
the gravitational wave detectors and balances could be suitable for detecting the Archimedes force. 
However, other aspects must be taken into account. The frequency bandwidth favored for detecting the effect at the periodical 
transition from normal to superconducting state obtained by a temperature change is in the region of 1-100 mHz, 
typical of the thermal processes. As a consequence the use of balances is the more suitable choice for this application.

\subsection{The balance and its seismic isolation}
A balance capable of measuring forces of the order of $\sim10^{-16} N$ requires an excellent isolation from the seismic 
noise at the very low frequency regime of 1-100 mHz. The use of a seismic-attenuation device based on the use of an inverted 
pendulum and a suitably tuned resonator hung to it (see fig. fig. \ref{fig:Balance}) can fullfil this requirement. 
The read-out system can be an optical lever or an interferometric sensor.
A challenging point is the design of the balance whose center of mass must be positioned on its bending point in order not 
to reinject the residual seismic noise in the system. 
The major sources of noise in this scheme are represented by the thermal
noise and the seismic noise.  As reported in fig. \ref{fig:noise}, for an integration
time of several months, this would permit to reach a good signal-to-noise
ratio.

\subsection{Thermal modulation study}
Another challenging point is related to the modulation of
the superconductor's temperature. Only the radiative mechanism must be used to remove or add thermal energy to the sample 
as it must be isolated from any external interaction that could add energy other than the vacuum one. The main difficulties 
are related to heating samples with mass of hundreds of $g$ within thermal transition time scales of hundreds seconds. 
To reduce the transition times and increase the frequency bandwidth the choice of the sample shape and of the heating system 
is crucial. With the help of a finite element analysis we have obtained a temperature modulation of about 1 K on a 150 mm 
diameter 0.5 mm thick disk. The shape is compatible with the present YBCO technology and is usefull to have 
signal within the expected sensitivity.

\section{Conclusion}
We have presented the Archimedes experiment aiming to verify the feasibility to measure the weigh of the vacuum. 
Three crucial points are focused with this project, the construction of a balance suitable to measure forces at the level 
of $10^{-16} ~ N$, a thermal modulation system using only a radiative heat exchange mechanism and the detailed study of 
high $T_c$ superconductive systems as the main source of vacuum energy.

\end{document}